# Half-metallicity of $Mn_2VAl$ ferrimagnet revealed by resonant inelastic soft x-ray scattering under magnetic field

R.Y. Umetsu[1,2,*], H. Fujiwara[3], K. Nagai[3], Y. Nakatani[3], M. Kawada[3], A. Sekiyama[3,4], F. Kuroda[5,6], H. Fujii[5,6], T. Oguchi[4,5,6], Y. Harada[7,8], J. Miyawaki[7,8], and S. Suga[5,9]

[1] Institute for Materials Research, Tohoku University, 2-1-1 Katahira, Sendai 980-8577, Japan

[2] Center for Spintronics Research Network, Tohoku University, 2-1-1 Katahira, Sendai 980-8577, Japan

[3] Division of Materials Physics, Graduate School of Engineering Science, Osaka University, 1-3 Machikaneyama, Toyonaka, Osaka 560-8531, Japan

[4] Center for Spintronics Research Network, Osaka University, 1-3 Machikaneyama, Toyonaka, Osaka 560-8531, Japan

[5] Institute of Scientific and Industrial Research, Osaka University, 8-1 Mihogaoka, Ibaraki 567-0047, Japan

[6] CMI²-MaDIS, National Institute for Materials Science, 1-2-1 Sengen, Tsukuba, Ibaraki 305-0047, Japan

[7] Institute for Solid State Physics, The University of Tokyo, Kashiwanoha, Chiba 277-8581, Japan

[8] Synchrotron Radiation Research Organization, The University of Tokyo, Sayo-cho, Hyogo 679-5198, Japan

[9] Forschungszentrum Jülich, PGI-6, 52425 Jülich, Germany

**Abstract**

Detailed information on the electronic states of both V and Mn $3d$ electrons in the ferrimagnet $Mn_2VAl$ is obtained by the bulk sensitive resonant inelastic soft x-ray scattering (SX-RIXS) excited with the circularly polarized light under an external magnetic field for the first time. The results under the V $L$-edge excitation have revealed the negligible partial density of states (PDOS) of the V $3d$ states around the Fermi energy as well as their rather localized character. Under the Mn $L$-edge excitation, on the other hand, the spectra are dominated by fluorescence with clear magnetic circular dichroism with noticeable excitation photon energy dependence. Compared with the theoretical prediction of the RIXS spectra based on the density-functional-theory band structure calculation, an itinerant, spin-dependent character of the Mn $3d$ states and decays of the Mn $2p$ core states are confirmed in consistence with the half-metallicity of the Mn $3d$ states.

Corresponding author: * rieume@imr.tohoku.ac.jp

## I. INTRODUCTION

Since the half-metallic electronic structure was predicted in the half-Heusler alloys of NiMnSb and PtMnSb [1] and in full-Heusler alloys such as Co-based Heusler alloys [2,3], a large number of investigations have been carried out from the interest in the field of spintronics. When electrons around the Fermi energy ($E_F$) are completely spin-polarized, a system must be very useful as a ferromagnetic electrode for the spin injection and tunnel magnetoresistance as well as various spin utilizable devices. For example, magnetoresistance of the magnetic tunneling junction with using complete half-metal ferromagnets would ideally be infinity. Very recently, other types of Heulser alloys such as Mn-based Heusler alloys and their quasi-ternary alloys have also been pointed out to show the half-metallic electronic states [4-10]. Among them $Mn_2VAl$ is one of the most attractive materials for device application. Its magnetic properties and theoretically predicted electronic structures were reported in early 1980's [11-13].

If the $Mn_2VAl$ orders completely, atoms of Mn, V and Al occupy the Wyckoff positions 8$c$, 4$b$, and 4$a$, respectively, with the space group $Fm\bar{3}m$ as schematically shown in Fig. S1 in Supplemental Material. The spontaneous magnetization per formula unit of 1.9 $\mu_B$/f.u. at 4.2 K is close to 2 $\mu_B$/f.u.[11] predicted by the generalized Slater-Pauling rule [14] and is much smaller than those in Co-based Heuser alloys. The magnetic critical Curie temperature $T_C$ of this ferrimagnet $Mn_2VAl$ is quite high to be about 760 K with antiferromagnetically coupled V and Mn spins [11,12]. This material is thought to be very promising for spintronic devices at room temperature because the expected current to switch its spin would be rather low. Investigations of $Mn_2VAl$ have recently been extensively carried out, for example, by the x-ray absorption magnetic circular dichroism (XAS-MCD) for bulk and film specimens [15-18].

For the fundamental investigations of half-metallic materials, researchers should pay attention how to provide convincing evidence of the half-metallic electronic states. To date fundamental magnetic properties reflecting the specific character of the half-metallic electronic states have been investigated. For example, rather small high-field magnetic susceptibility [19] and the negligibly small pressure dependence of the magnetization in some Co-based Heusler alloys were reported. These results are understood as reflecting the fact that the electronic state near $E_F$ is insensitive to applied external fields in the case of half-metallic ferromagnets [20]. Recently, anisotropic magnetoresistance is also predicted to serve as one of the practical screening tests of the half-metallic ferromagnets from both theoretical and experimental aspects [21,22].

We report here for the first time the direct evidence of the spin-polarized electronic structures of the half-metallic ferrimagnet Mn$_2$VAl Heusler alloy with highly ordered $L2_1$-type structure by means of resonant inelastic soft x-ray scattering (SX-RIXS) measurements of the V and Mn 2$p$ core excitation with use of right and left helicity circularly polarized light and with external magnetic field. RIXS is a bulk sensitive photon-in and photon-out spectroscopy, and very powerful to investigate such as $d$-$d$ excitations for open shell 3$d$ orbitals and magnetic excitations for spin systems [23-26] as well as 2$p$-3$d$ transitions in element- and symmetry-specific ways. These excitations and decays are sensitive to spin, electron correlation, crystalline symmetry, and the strength of hybridization with the ligand band. Furthermore, RIXS is insensitive to the surface conditions because of its long probing depth (> 100 nm) in contrast to any kind of photoelectron spectroscopy (PES). Since the emitted light is probed, RIXS is not affected by such an external perturbation as magnetic field in contrast to PES for electrons. In the present study, the photon energy ($h\nu_{in}$) dependence of the magnetic circular dichroism (MCD) of RIXS was measured in detail in order to obtain the spin-dependent information. The results obtained in the present experiments and theoretical analyses confirmed the half-metallicity of Mn$_2$VAl, demonstrating that the RIXS and RIXS-MCD are extremely powerful for the study of the electronic structures of the half-metallic ferromagnetic or ferrimagnetic materials.

## II. EXPERIMENTAL
### A. Sample preparation

Mother ingot of polycrystalline Mn$_2$VAl was fabricated by induction melting in an argon atmosphere. Since the vapor pressure of Mn is high during the melting, excess Mn elements are contained in the mother ingot. Single crystal was grown by the Bridgeman method with a size of 12 mm in diameter and about 30 mm in length. The obtained ingot was annealed at 1473 K to grow the crystal grains. Furthermore, a two-step annealing process at 1123 K and then 873 K was employed in order to control the microstructures and to heighten the degree of order. These sample preparation processes resulted in rather high degree of order as $S$ = 0.84 in our sample compared with $S$ = 0.5 by Kubota *et al*. [15] and $S$ = 0.4 by Meinert *et al*. [17] in their samples. Crystal orientation was checked by the back Laue method and the specimen was cut in a strip form in the direction parallel to <100>. The sample composition was confirmed to be Mn: 50.5, V: 26.9, Al: 22.6 (atomic %) with an electron probe microanalyzer. Sample magnetization was measured with a superconducting quantum interference devices (SQUID) magnetometer. Magnetization (*M-H*) curve measured at 5 K for Mn$_2$VAl showed that more that 90% of saturation magnetization is realized already at 0.2 T of the external magnetic field. The expected value of the magnetic moment for $Mn_2VAl$ by Slater-Pauling rule, which is predicted by Galanakis *et al*. [14] is 2 $\mu_B$/f.u. Slight deviation of the saturation magnetic moment 1.82 $\mu_B$/f.u. of our sample from 2 $\mu_B$/f.u. might be caused by a small amount of the off-stoichiometric composition. Although atomic content of Mn is well controlled in the present specimen, atomic content of V is slightly increased compared to that of Al.

### B. Measurements

Resonant inelastic soft x-ray scattering (SX-RIXS) for V and Mn $2p$ core excitation was measured at room temperature at the HORNET end station installed in the long undulator beam line BL07LSU of SPring-8, Japan [27,28]. Measurement was performed with an external magnetic field of 0.25 T, which was applied by a permanent magnet with two poles for passing the excitation light [29,30]. The direction of the magnetic field was repeatedly reversed by the rotatable feed through supporting the magnet (The schematic view of the experimental geometry is given in Fig. S2 in Supplemental Material). Right and left helicity circularly polarized lights (RCP and LCP) parallel to the magnetic field were incident at 45° onto the (100) plane for excitation. Light emitted at 45° from this surface was dispersed by a grating and detected by a two dimensional detector. The <100> axis was contained in the scattering plane. The direction of the magnetic field was reversed to confirm the genuine magnetic circular dichroism in this RIXS experiment. The total energy resolution was set to ~140 (170) meV at the V (Mn) $2p_{3/2}$ edge. RIXS measurements were performed on a surface of the sample obtained by fracturing in an Ar globe box in advance and transferred to a RIXS chamber with a vacuum of $1\times10^{-5}$ Pa without any exposure to atmosphere.

### III. RESULTS AND DISCUSSION
### A. RIXS and RIXS-MCD experimental spectra

Figures 1(a) and 1(b) show the x-ray absorption spectra (XAS) of V and Mn-$L_3$ edges, respectively, by means of the total electron yield recorded at 20 K [18]. The numbers above the vertical bars on the XAS indicate the incidence photon energy $h\nu_{in}$ for RIXS spectra. Figures 1(c) and 1(d) show the RIXS spectra for the V and the Mn $L_3$-edges at room temperature, respectively, measured by parallel ($\mu^+$: blue) and antiparallel ($\mu^-$: red) configurations between the light helicity and the direction of the magnetic field as a function of the energy loss given by the horizontal axis. In the figures, strong intensity peaks without any energy loss are always observed and they are called the elastic component. In the larger energy loss region, the so-called fluorescence peaks are observed with their energy loss increasing linearly with $h\nu_{in}$. The relative height of the fluorescence peak is noticeably smaller than that of the elastic peak in the case of V $L_3$-edge excitation. The difference of $\mu^+$ and $\mu^-$ is clearly observed in Mn $L_3$-edge excitation.

Figures 2(a) and 2(b) show the intensity maps of the RIXS spectra for the V $L_3$-edge obtained by Fig. 1(c). The horizontal axis shows the energy loss from the incident photon energy $h\nu_{in}$. The spectral differences of the RIXS between the parallel and antiparallel configurations, RIXS-MCD, are shown in Fig. 2(c). The corresponding results for the Mn $L_3$-edge are shown in Figs. 2(d)-2(f), respectively. From the comparison of these spectra between the V and the Mn $L_3$-edges, several significant features are recognized. The most characteristic feature is that the fluorescence peak associated with the V $L_3$-edge does not branch off from the elastic peak. On the other hand, there is almost no gap in the energy loss between the elastic peak and the appearance of the fluorescence peak for the Mn $L_3$-edge. In addition, weak structure is observed around the constant energy loss ~2 eV for a wide excitation region above $h\nu_{in}$ ~515 eV for V $L_3$-edge. This inelastic energy loss feature is considered to be associated with the $d$-$d$ excitation because the existence at a constant energy loss cannot be due to any fluorescence feature. If we compare the fluorescence MCD for the V and Mn $L_3$-edges in Figs. 2(c) and 2(f), it is also recognized as the negative Mn fluorescence MCD is spread in the photon energy range in $h\nu_{in}$ = 638 - 639 eV, while very weak positive V fluorescence MCD is observed around $h\nu_{in}$ = 513 eV.

Typical RIXS spectra and its MCD at $h\nu_{in}$ = 512.5 eV for the V $L_3$-edge and $h\nu_{in}$ = 638.6 eV for the Mn $L_3$-edge are reproduced in Figs. 3(a) and 3(b), respectively, by solid lines. The RIXS-MCD features are observed in both cases of V and Mn $L_3$-edge excitations, and the sign of the MCD of the major fluorescence feature is found to be opposite between the cases of V and Mn in agreement with the ferrimagnetic character of this material, $Mn_2VAl$. Here, one notices that a broad peak is observed in the V $L_3$-edge excitation, while double peak features split by 1.0 ~ 1.2 eV are observed in Fig. 3(b) in the case of the Mn $L_3$-edge excitation beside the elastic peak. Detailed RIXS-MCD for the Mn $L_3$ threshold excitation with changing $h\nu_{in}$ is shown later in Figs. 4(a1)-4(a6). These characteristic features must be closely correlated with the electronic structures of V and Mn.

### B. Theoretical basis

For interpreting the observed RIXS and MCD spectra, theoretical calculations were performed by means of the density functional theory (DFT). Based on the DFT, spin resolved partial density of states (PDOS) of the V, Mn and Al are calculated as shown in Fig. 3(c) together with the total density of states (DOSs). The $e_g$ and $t_{2g}$ derived components are separately shown in Figs. 3(d) and 3(e) for V and Mn, respectively. First of all, the PDOS of Al is almost negligible near $E_F$. In the case of the V 3$d$ states, the PDOS are found to be very small in the region of $E_F \pm 0.6$ eV. High PDOS of V of the $t_{2g}$ occupied states are around -1.5 eV and those of the unoccupied $t_{2g}$ and $e_g$ states are located at around +1.6 eV. In the case of the Mn 3$d$ states, however, rather complex electronic structures strongly dependent on the spin are predicted. Although the occupied valence bands below -0.4 eV are composed of both $t_{2g}$ and $e_g$ states with both spin up and down states, clear differences are recognized above this energy. Namely, spin down $t_{2g}$ states are crossing $E_F$ with high PDOS and the spin up $t_{2g}$ states have negligible PDOS between -0.5 eV and +0.7 eV. In the case of the Mn $e_g$ states, on the other hand, up spin states have negligible PDOS between -0.4 and +0.4 eV, though down spin states have a certain PDOS between -0.4 and +1 eV before showing high PDOS around +1.4 eV. Thus half metallic PDOS behavior so far predicted is reconfirmed in Fig. 3(e).

As already pointed out in Figs. 2(a) and 2(b), the fluorescence component of RIXS in V $L_3$-edge does not branch off from the elastic peak. The absence of any additional fluorescence peak between the elastic peak and the peak around ~2 eV for the $L_3$ threshold excitation is consistent with the weakness or negligible PDOSs of the V 3$d$ states around $E_F$ for both spin up and spin down V 3$d$ states. The energy splitting between the V unoccupied $e_g+t_{2g}$ states PDOS and the occupied $t_{2g}$ PDOS ranges from 2 to 4 eV. Although this predicted $d$-$d$ splitting energy is slightly larger than the experimental $d$-$d$ splitting energy roughly estimated as 2-3 eV in Fig. 3(d), the constant loss energy feature in Figs. 2(a) and 2(b), is unambiguously ascribed to the genuine RIXS feature due to the $d$-$d$ excitation. On the other hand, RIXS feature for the Mn $L_3$-edge excitation branches off really from the elastic peak and moves linearly with $h\nu_{in}$ revealing its fluorescence origin, suggesting the finite PDOS at $E_F$ of the Mn 3$d$ states. In the PDOS for the Mn shown in the Fig. 3(e), finite PDOS exists around the $E_F$ in the down spin state, supporting the experimental results.

The RIXS-fluorescence spectra were simulated based on the Kramers-Heisenberg formula as shown in equation (8) in the Supplemental Material [31], where details of the calculations are explained. Since the correlation between the fluorescence RIXS-MCD and the band structures must be discussed, more detailed calculation with taking the magnetization into account must be performed. Under the external magnetic field of 0.25 T, magnetization is almost saturated with the magnetic moment along the magnetic field [11,18]. Simulated RIXS-fluorescence spectra and the MCD based on the DFT for V and Mn $L_3$-edges are shown by dotted lines in the Figs. 3(a) and 3(b), respectively, in addition to the experimental spectra. It is seen that the simulated RIXS-MCD qualitatively reproduces the experimental feature, for example, such as the double peak feature of the fluorescence in the Mn $L_3$-edge. If we compare the spectra of theoretically predicted fluorescence and its MCD with corresponding experimental results, one notices that the predicted spectra are more widely energy spread than the experimental results. This may be due to the larger electron correlation energy between the Mn $3d$ electrons beyond that employed in the ordinary DFT calculation.

### C. Detection of the spin polarized Mn $3d$ electronic states

In order to further discuss the Mn $3d$ states around $E_F$, RIXS and the MCD were systematically investigated in the Mn $L_3$-edge threshold excitation region. Figures 4(a1)-4(a6) show the experimental RIXS spectra for $\mu^+$ and $\mu^-$ configurations and their MCD with changing the $h\nu_{in}$ (638.2 ≤ $h\nu_{in}$ ≤ 639.2 eV). Figure 5 shows schematic views of the Mn $L_3$ RIXS-MCD processes by considering the PDOS for the half-metallic magnetic system. The difference of the photon energy $h\nu_{in}$ from the energy $E_{2p0}$ between the $E_F$ and the $m_j = -3/2$ Mn $2p$ core state is defined hereafter by $h\nu_1$. PDOS of the up (down) spin state is schematically shown on the left (right) hand side of the energy axis (vertical axis) in each figure. The Zeeman splitting of the Mn $2p$ core level states of around 0.5 eV due to the effective magnetic field induced by the spin-polarized $3d$ states is taken into account in the present simulation. The excitation to the empty conduction band states is shown in Fig. 5 by the upward arrows with the filled circles with assuming the gap of 0.3 eV from the $E_F$ to the bottom of the DOS in the majority up spin state of the Mn $3d$ states.

Five energy ranges of $h\nu_1 = h\nu_{in} - E_{2p0}$ can be considered for the excitation as 1) below 0.16 eV, 2) 0.16 − 0.46 eV, 3) 0.46 − 0.63 eV, 4) 0.63 − 0.8 eV and 5) above 0.8 eV. Since the experimental results were obtained at 300 K, one should note that the energy broadening of around 0.1 eV cannot be neglected in the comparison between the experimental results and theoretical prediction. In both the $2p$ core excitation and the $3d$-$2p$ fluorescence decay processes, the spin is conserved in the dipole transition. However, the hole spin can be relaxed before the fluorescence decay in the core hole with $m_j = -1/2$ and $1/2$ states which are composed of both spin up and down states due to the spin-orbit coupling. So even when the $m_j = -1/2$ state with spin down state is excited to the empty conduction band, the core hole spin can be partially relaxed to the spin up state before the fluorescence takes place. This means that one should take into account the fluorescence in both spin down and up channels with the fixed relative weight given in the initial core hole states when the fluorescence decay to the core holes with $m_j = -1/2$ or $1/2$ states are calculated. In Fig. 5, the width of the arrows corresponds to the transition probability.

The fluorescence spectra for the Mn $L_3$-edge excitation are calculated as shown in Figs. 4(b1)-4(b6). In the experimental MCD spectra in Figs. 4(a1)-4(a6), the intensity increases with increasing $h\nu_{in}$ showing the double peak future, which becomes less clear above $h\nu_{in}$ = 638.8 eV. Such a tendency is qualitatively reproduced by the calculated spectra. Figures 4(c1)-4(c6) show the predicted MCD spectra of fluorescence to the Mn 2p $m_j$ = -3/2, -1/2, 1/2 and 3/2 states. It is clear that the fluorescence decay to the $m_j$ = -3/2 state is dominating at the low $h\nu$ threshold. Since the down spin $t_{2g}$ states have substantial PDOS of the unoccupied states from $E_F$ to +0.8 eV, and the down spin $e_g$ states have high PDOS around +1.4 eV, the pure down spin $m_j$ = -3/2 states can be continuously excited at least up to $h\nu_1 \sim$ 0.8 eV in addition to the region in +1.2 - 1.6 eV inducing the remarkable negative fluorescence MCD in Figs. 4(c1)-4(c6). Since the PDOS of the occupied Mn 3d states in the region from $E_F$ down to -0.4 eV for the down spin state is rather high, noticeable fluorescence and its clear MCD is predicted just from the energy loss of 0 eV as clearly seen in Figs. 4(b1) and 4(c1), 4(b2) and 4(c2). The fluorescence feature with an energy loss peak near 0.3 eV in Figs. 4(b1) and 4(b2) is thought to reflect the high PDOS of the down spin occupied $t_{2g}$ states in the above mentioned region.

With increasing $h\nu_1$ up to ~0.46 eV, excitation from the $m_j$ = -1/2 state becomes gradually feasible in addition to the excitation from the $m_j$ = -3/2 state, providing negative but small MCD of the fluorescence to the $m_j$ = -1/2 state as shown in Figs. 4(c2) and 4(c3). The origin of the doublet feature of the fluorescence and its MCD separated by ~1 eV (Figs. 4(a2), 4(a3) and 4(a4)) is relatively well predicted by the calculation as shown in Figs. 4(b1)-4(b4) and 4(c1)-4(c4). Both of the doublet features must be due to the fluorescence transition into the down spin $m_j$ = -3/2 core hole state. Judging from the sum of the Mn $t_{2g}$ and $e_g$ down spin states shown in Fig. 3(e), which is equal to the Mn down spin PDOS in Fig. 3(c), two peaks of the occupied down spin Mn 3d PDOS are recognizable at ~0.8 and ~1.8 eV. These two peaks are surely contributing to the doublet fluorescence features of Fig. 3(b) as well as Figs. 4(a2)-4(a4).

For 0.46 eV < $h\nu_1$ < 0.63 eV, both down and up spin states are excited from the $m_j$ = -1/2 core states. For 0.63 eV < $h\nu_1$ < 0.8 eV, the similar situation takes place for the $m_j$ = +1/2 core states. In each case, both spin down and up states are excited and fluorescence takes place after the relaxation of the core hole spin. With the increase in the $h\nu_1$, the relative weight of the transition to the $m_j$ = -1/2, +1/2 and +3/2 states increase gradually with the increase of the magnitude of the individual fluorescence MCD. As a result, the total magnitude of the predicted fluorescence MCD decreases relatively as seen in the series of Figs. 4(b1) to 4(b6) in consistence with the experimental results in Fig. 4(a3) to 4(a6).

Various investigations to directly check the half-metallic electronic structure have been performed up to now. Ultraviolet-photoemission spectroscopy (UPS) and hard x-ray angle resolved photoelectron spectroscopy (HAXARPES) [32-36] are such examples. In the case of UPS, it is known that the surface electronic structure accessible by UPS is noticeably different from that in the bulk. The HAXARPES is much more bulk sensitive. However, its orders of magnitude reduced photoionization cross sections for the valence band electronic states compared with the UPS strongly hinder the usefulness of HAXARPES studies. Moreover the possible recoil shift effects for the valence electron states spoil reliable discussions on the most important electronic structures near the $E_F$ in Mn$_2$VAl [32,35,36]. Although spin-polarized and angle-resolved photoelectron spectroscopy (SP-ARPES) will be desired to detect the half-metallicity, its low detection efficiency (orders of $10^{-4}$ or less) makes it almost impossible to make reliable experiment. Therefore, the electronic structure, especially half-metallicity of the bulk Heusler alloys has not yet been fully clarified. In the present study, detailed $h\nu_{in}$ dependence of RIXS-MCD was clearly observed, providing the spin-dependent information. The qualitative agreement obtained in the present experiments and theoretical analyses confirmed the half-metallicity of Mn$_2$VAl, demonstrating that the RIXS and RIXS-MCD are extremely powerful for the study of the electronic structures of the half-metallic ferromagnetic or ferrimagnetic materials.

## IV. CONCLUSION

Resonant inelastic x-ray scattering (RIXS) measurement in magnetic field was performed on the single crystal half-metallic Heusler alloy, Mn$_2$VAl, for the first time in order to obtain reliable information on the bulk electronic state of the 3$d$ electrons. The $d$-$d$ excitation due to the $t_{2g}$-$e_g$ splitting is clearly observed for V in Mn$_2$VAl under the $L_3$-edge excitation. The loss energy of the V $d$-$d$ RIXS maximum is found to be about 2 eV, being comparable to the splitting energy between the theoretically predicted $e_g$ and $t_{2g}$ states. The delayed branching off in the V 3$d$-2$p$ fluorescence peak from the elastic peak demonstrates the nearly absent V 3$d$ PDOS around $E_F$. The clear appearance of the $t_{2g}$-$e_g$ RIXS of V reflect the rather localized character of the V 3$d$ states. The RIXS-MCD of the fluorescence peaks of Mn 3$d$-2$p$ transition under the $L_3$-edge excitation shows negative sign with clear $h\nu_{in}$ dependence. The sign and the shape of the RIXS-MCD are qualitatively reproduced in consistence with the DFT calculations and confirmed the absence of the up spin Mn 3$d$ PDOS at the $E_F$, demonstrating the half-metallicity of Mn$_2$VAl Heusler alloy. Thus the bulk sensitive RIXS studies under external magnetic field are confirmed to be essential to study the detailed electronic structures of various Heusler alloys and family materials.

## ACKNOWLEDGMENTS

We sincerely thank K. Yubuta, T. Sugawara and Y. Murakami for their helps to make single crystals and EPMA experiments, and Prof. T. Kanomata for valuable discussions. This research was supported by 1)Precursory Research for Embryonic Science and Technology (PRESTO), Japan, 2)Science and Technology Agency (JST), Japan and 3)Grant-in-Aid for Challenging Exploratory Research from the Japan Society for the Promotion of Science (JSPS). This work was carried out under the approval of BL07LSU at SPring-8 Synchrotron Radiation Research Organization and the Institute for Solid State Physics, The University of Tokyo (Proposal No. 2016BG05, 2016B7512). Partial fundamental measurements were carried out at the Center for Low Temperature Science, Institute for Materials Research, Tohoku University.

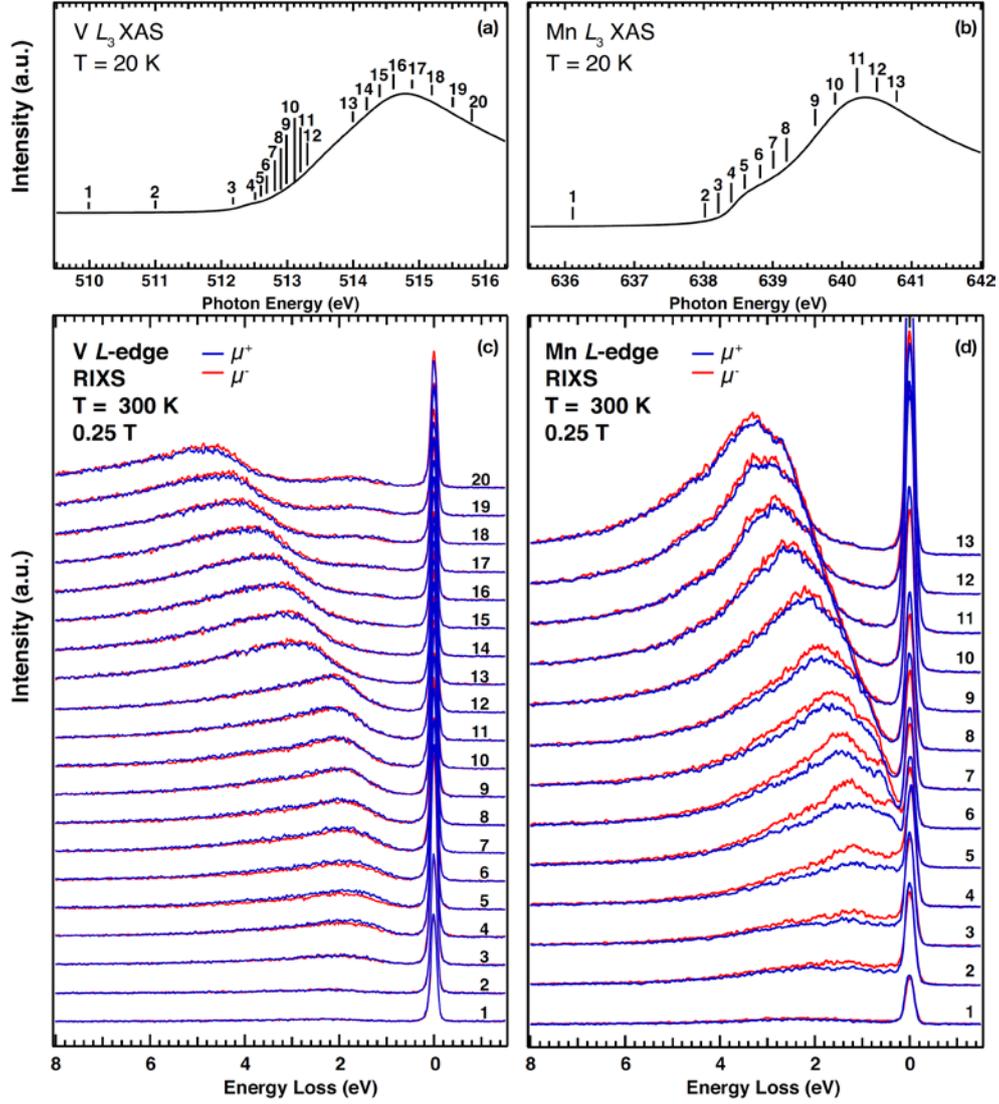

FIG. 1 XAS and RIXS spectra for V and Mn $L_3$-edges. (a) and (b) are XAS for the V and the Mn $L_3$-edges, respectively, at 20 K and 2 T magnetic field [18]. (c) and (d) are the RIXS spectra for the V and the Mn $L_3$-edges, respectively, obtained at room temperature. In (c) and (d), the RIXS spectra were measured by parallel ($\mu^+$) and antiparallel ($\mu^-$) configurations between the light helicity and the direction of the magnetic field. The numbers above the vertical bars on the XAS indicate the excitation photon energy $h\nu_{in}$ for the RIXS spectra.

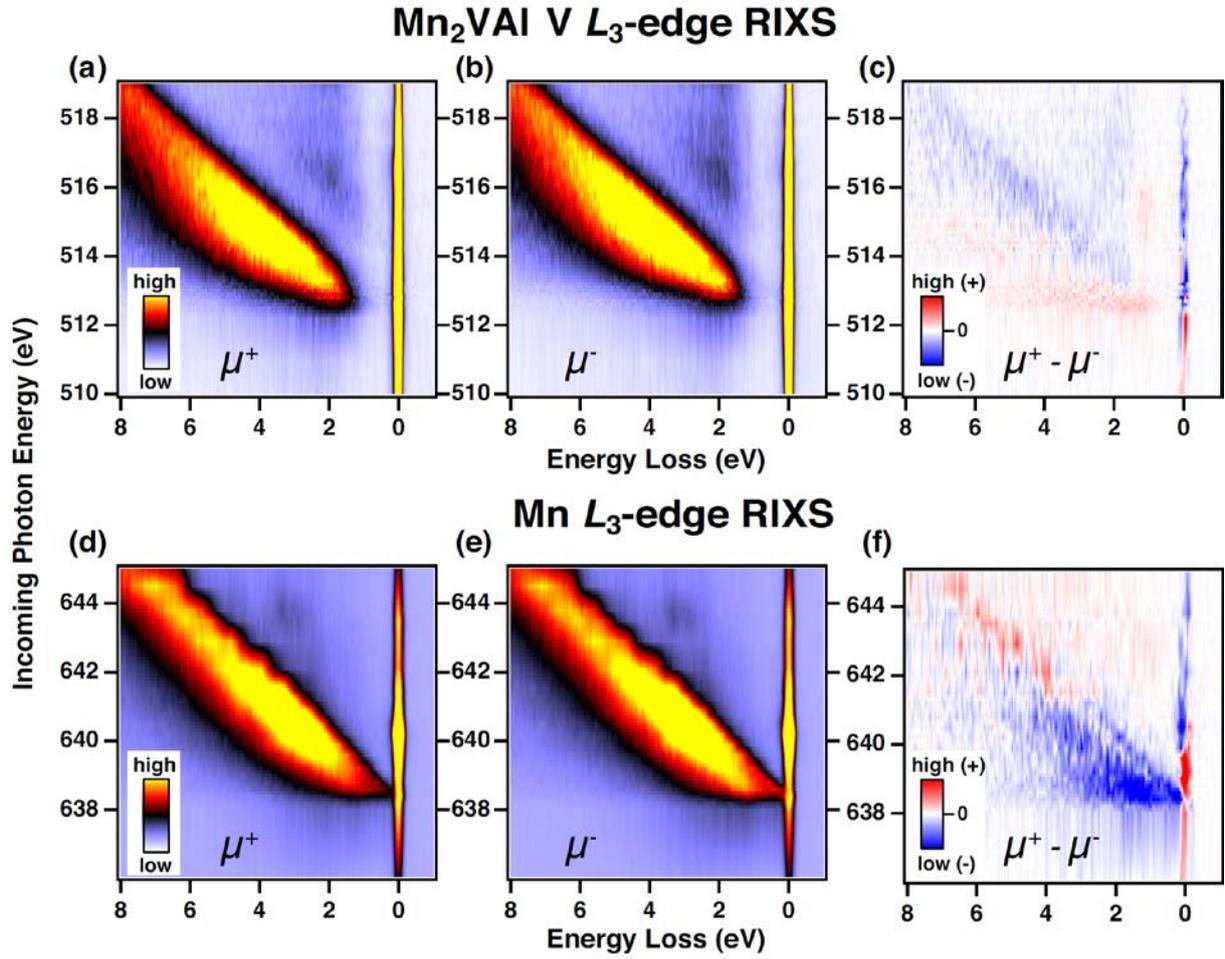

FIG. 2 Intensity maps of $h\nu_{in}$-dependent RIXS and MCD at V and Mn $L_3$-edges in Mn$_2$VAl. (a) and (b) are the intensity maps of the RIXS of V as a function of $h\nu_{in}$ obtained at room temperature in a magnetic field of 0.25 T for $\mu^+$ and $\mu^-$ configurations. RIXS MCD is given by $\mu^+$ - $\mu^-$ in (c). Figures (d), (e) and (f) are the corresponding results for the Mn $L_3$-edge.

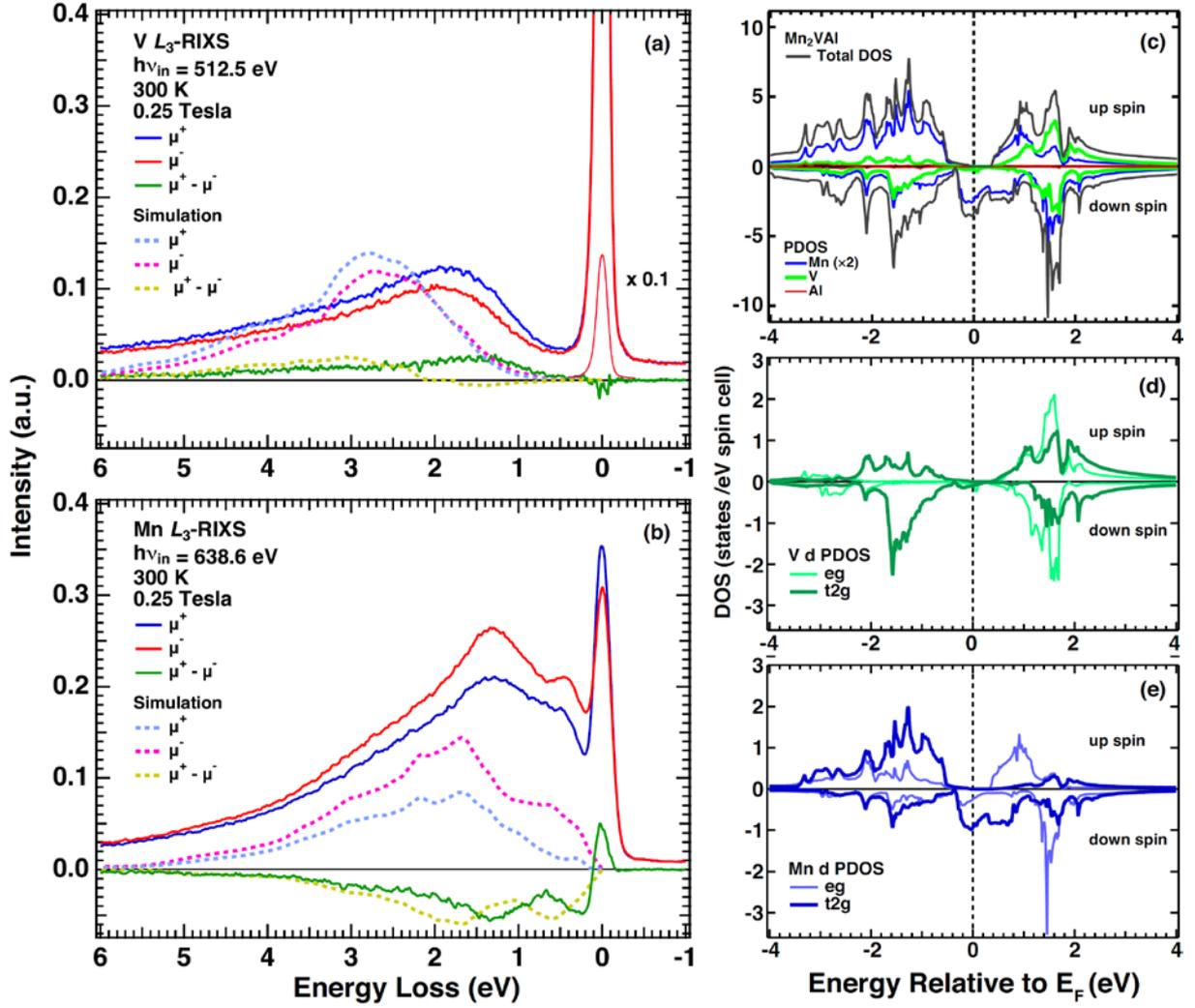

FIG. 3 Representative RIXS spectra for V and Mn $L_3$-edges, and theoretically predicted PDOSs of Mn, V and Al in $Mn_2VAl$. RIXS spectra recorded for $\mu^+$ and $\mu^-$ configurations at incoming photon energy, $h\nu_{in}$ of 512.5 eV (a) and 638.6 eV (b) at the V and Mn $L_3$-edges, respectively, together with the MCD. Figures (a) and (b) include simulated spectra based on the DFT for V and Mn $L_3$-edge, respectively. Figure (c) shows the theoretical prediction of the spin dependent total DOS as well as PDOSs of Mn, V and Al. Figures (d) and (e) correspond to the $e_g$ and $t_{2g}$ components of the partial DOSs of V and Mn, respectively.

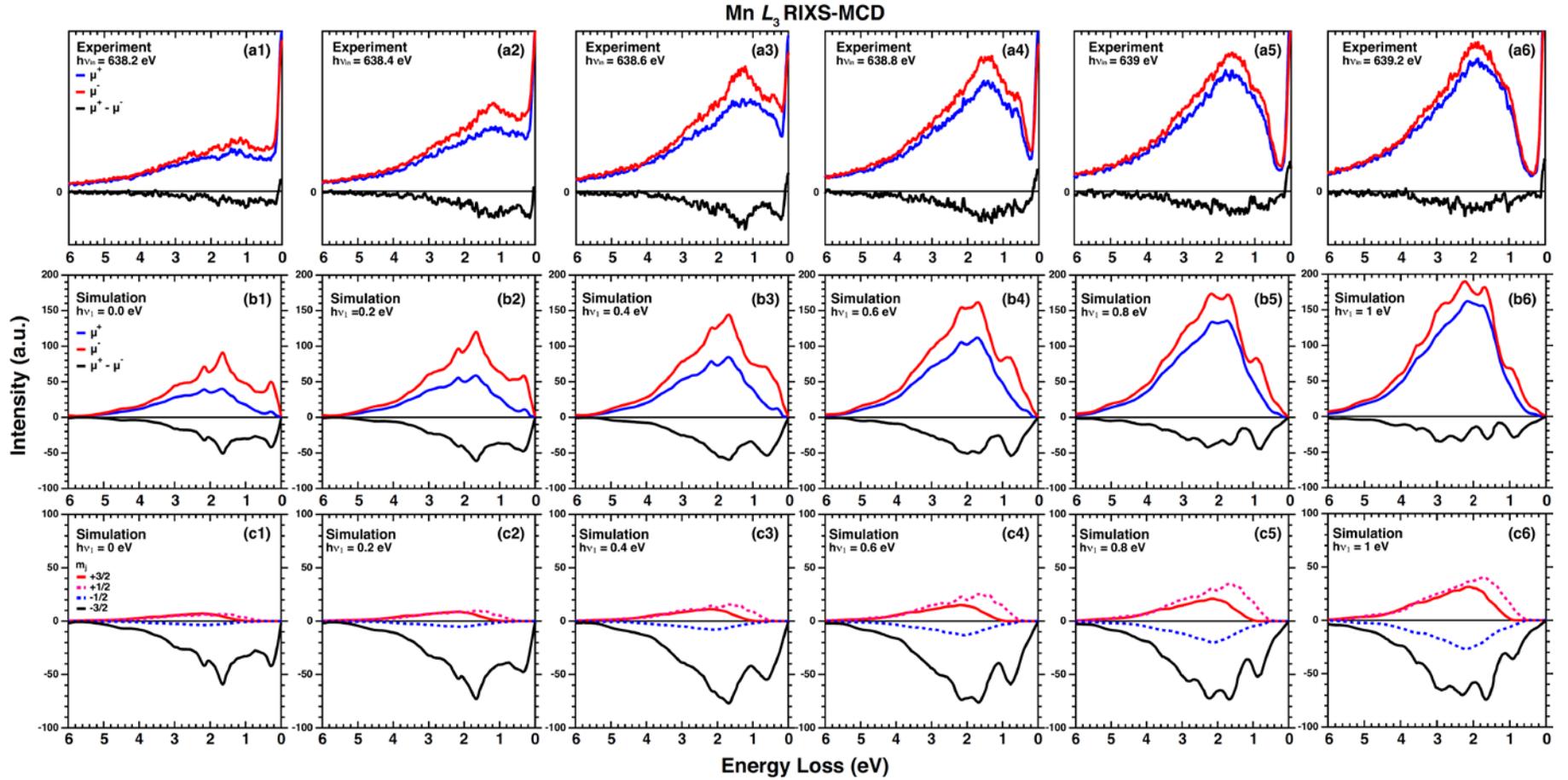

FIG. 4 RIXS and RIXS-MCD spectra for Mn $L_3$-edges. Figures (a1)-(a6) show the experimental RIXS results for the $\mu^+$ and $\mu^-$ configurations and their MCD spectra for the Mn $L_3$-edge as a function of the energy loss. Theoretically calculated RIXS and RIXS-MCD spectra based on DFT calculation are in (b1)-(b6). The definition of $h\nu_1$ is given in the text. Figures (c1)-(c6) show $m_j$ resolved RIXS-MCD spectra for Mn $L_3$-edge.

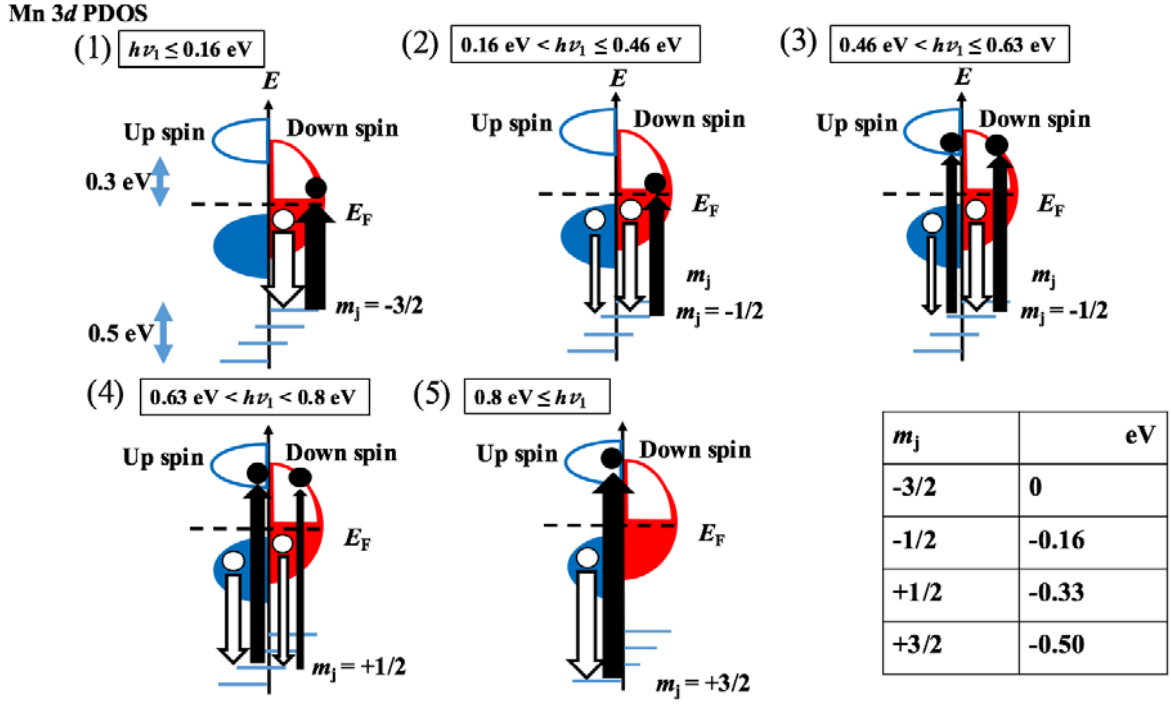

FIG. 5 Schematic views of the RIXS-MCD processes in the case of half-metallic density of states. Excitation from the Zeeman split 2p $m_j$ core states is considered in the present simulation as indicated in the table and the Zeeman splitting is assumed to be around 0.5 eV due to the effective magnetic field. The difference of the photon energy ($h\nu_{in}$) from the energy $E_{2p0}$ between the Fermi energy ($E_F$) and the Mn 2p $m_j = -3/2$ state is given here by $h\nu_1$. PDOS of the up (down) spins state is schematically shown on the left (right) hand side of the energy axis (vertical axis) in each figure. A gap of 0.3 eV from the $E_F$ to the bottom of the DOS in the majority up spin state is also assumed to exist. The spin is thought to be fully down and up in the $m_j = -3/2$ and 3/2 states, respectively. On the other hand, the spin at $m_j = -1/2$ and 1/2 states is composed of both spin up and down states due to the spin-orbit coupling. The width of the arrows corresponds to the transition probability.

# SUPPLEMENTAL MATERIAL
## SA. CRYSTAL STRUCTURE OF Mn$_2$VAl HEUSLER ALLOY

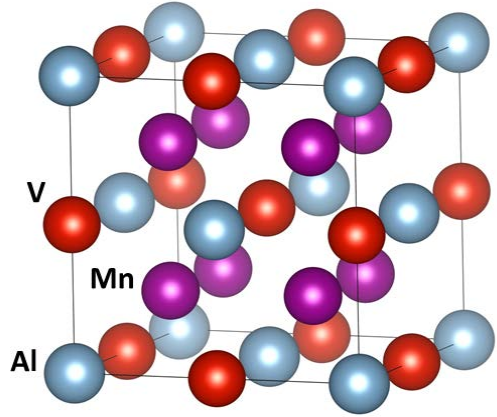

FIG. S1. Crystal structure of Mn$_2$VAl Heusler alloy with $L2_1$-type [1]. If the Mn$_2$VAl orders completely, Mn atoms (purple) occupy the Wyckoff position 8$c$, V atoms (red) 4$b$, and Al atoms (grey) 4$a$ with the space group $Fm\bar{3}m$.

## SB. EXPERIMENTAL GEOMETRY

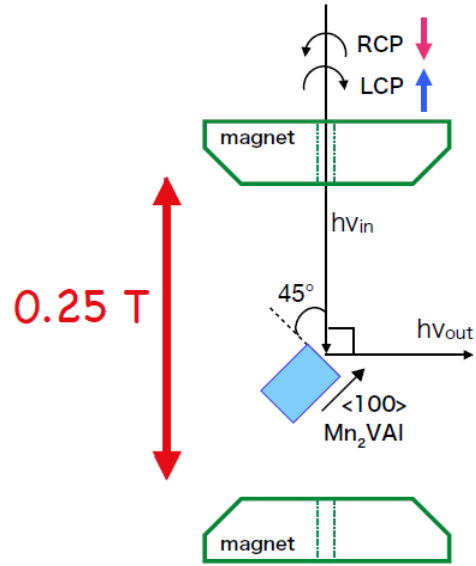

FIG. S2. Schematic view of the scattering geometry in SX-RIXS measurements. The single crystal of Mn$_2$VAl specimen was tilted 45° from the direction of the incident x-ray. Magnitude of the magnetic field in the central region of the yoke-type magnet is about 0.25 T, which is enough to saturate the magnetization for Mn$_2$VAl. The right and left helicity circularly polarized incident photon is used for RIXS excitation and the magnetic circular dichroism (MCD) is defined by the difference of the RIXS spectra for the $\mu^+$ and $\mu^-$ configurations.

## SC. SIMULATION BASED ON THE DENSITY FUNCTIONAL THEORY

The electronic structure calculation based on DFT has been performed using the HiLAPW code, which is based on the all-electron full-potential augmented plane-wave (FLAPW) method [2]. The generalized gradient approximation (GGA) using the Perdew-Burke-Ernzerhof scheme has been used for the exchange correlation potential [3,4]. The relativistic effects are considered for the 2p core states including the spin-orbit coupling. On the other hand, the spin-orbit coupling for the 3d states are negligible compared with the 2p states, since the orbital magnetic moment of Mn and V were estimated as 0.026 $\mu_B$/Mn and 0.037 $\mu_B$/V by XMCD measurements, respectively [5]. Plane-wave expansion cutoffs were set to 20 Ry for the wave functions and 80 Ry for the charge density and potential functions. The muffin-tin sphere radius was chosen as 1.1 Å for all elements. For the Brillouin-zone integration, a 16×16×16 uniform mesh was used with the tetrahedron integration technique. The atoms were placed on the general form $X_2YZ$ of the $L2_1$ structure with $X$ at 8c site in the Wyckoff position, $Y$ at 4b site, and $Z$ at 4a site. The lattice constant was set to 5.875 Å [6].

The RIXS spectra were simulated by using the Kramers-Heisenberg formula [7,8] described as

$$\sigma(\nu_{in}, \nu_{out}) \propto \sum_f |\sum_i \frac{\langle b|\hat{F}_2^{(\mu_2)}|i\rangle \langle i|\hat{F}_1^{(\mu_1)}|a\rangle}{E_i - E_a - h\nu_{in} - i\Gamma_i}|^2 \times \delta((h\nu_{in} - h\nu_{out}) - (E_b - E_a)) \quad (1)$$

where $a$, $b$, and $i$ denote the initial, final, and intermediate states having the energy of $E_a$, $E_b$ and $E_i$, respectively. The incoming and outgoing photon energies are described as $h\nu_{in}$ and $h\nu_{out}$, respectively. Lifetime broadening of a core hole is given by $\Gamma_i$ of 0.36 (0.28) eV for Mn (V) [9], and the dipole transition operator is described as $\hat{F}^{(\mu)}$ for the photon helicity ($\mu$) of the circularly polarized light. The electron configuration of the initial, intermediate and final states with $|a\rangle = |2p^6 v\rangle$, $|i\rangle = |2p^5 ve\rangle$, and $|b\rangle = |2p^6 v^{-1} e'\rangle$ are taken into account with the relative energy of $E_a = 0$, $E_i = \varepsilon_e - E_{2p}$, and $E_b = \varepsilon_{e'} - \varepsilon_{v^{-1}}$ [10]. Here $v$ and $e$ represent the valence electrons and the electron in the empty conduction band, and $v^{-1}$ and $\varepsilon_{v^{-1}}$ stand for the valence electron state with one hole induced by the decay into the 2p core hole state and the energy of valence electrons with one hole. Then the denominator of equation (1) can be expressed by considering these energies and the content of the δ function as

$$E_i - E_a - h\nu_{in} - i\Gamma_i = \varepsilon_{v^{-1}} + \varepsilon_e - \varepsilon_{e'} - E_{2p} - h\nu_{out} - i\Gamma_i$$

If we assume $\varepsilon_e = \varepsilon_{e'}$, where no change is considered for the electron excited into the empty conduction band on the fluorescence decay, the denominator of equation (1) is further approximated as

$$E_i - E_a - h\nu_{in} - i\Gamma_i = \varepsilon_{v^{-1}} - E_{2p} - h\nu_{out} - i\Gamma_i \tag{2}$$

On the same assumption the transition matrix elements are transformed as

$$\langle i|\hat{F}_1^{(\mu_1)}|a\rangle = \langle e|\hat{F}_1^{(\mu_1)}|2p\rangle \tag{3}$$

$$\langle b|\hat{F}_2^{(\mu_2)}|i\rangle = \langle v^{-1}|\hat{F}_2^{(\mu_2)}|2p^{-1}\rangle \delta(e'-e) \tag{4}$$

The summation for the intermediate states $|i\rangle$ can be expressed by the two integrals as follows and the RIXS intensity can be described as,

$$\begin{aligned}\sigma(\nu_{in},\nu_{out}) \propto \sum_f \int d\varepsilon_{v^{-1}} \int d\varepsilon_e \\ \times \frac{|\langle v^{-1}|\hat{F}_2^{(\mu_2)}|2p^{-1}\rangle|^2 \cdot |\langle e|\hat{F}_1^{(\mu_1)}|2p\rangle|^2}{(\varepsilon_{v^{-1}} - E_{2p} - h\nu_{out})^2 + \Gamma_i^2} \\ \times \delta((h\nu_{in} - h\nu_{out}) - (\varepsilon_e - \varepsilon_{v^{-1}}))\end{aligned} \tag{5}$$

If we assume that the square of the matrix element in the numerator is proportional to the partial density of states of the occupied (unoccupied) states denoted as $D^{occ}$ ($D^{unocc}$) at the energy of $\varepsilon_{v^{-1}}$ ($\varepsilon_e$) of occupied (unoccupied) valence states multiplied by the transition probability $w_{jm_jm_s}^{(\mu)}$ between the 2p and 3d states depending on the helicity ($\mu$) [10,11,12], we obtain

$$|\langle e|\hat{F}_1^{(\mu_1)}|2p\rangle|^2 \propto \sum_{jm_jm_s} w_{jm_jm_s}^{(\mu_1)} D_{(m+\mu_1)m_s}^{unocc}(\varepsilon_e) \tag{6}$$

$$|\langle v^{-1}|\hat{F}_2^{(\mu_2)}|2p^{-1}\rangle|^2 \propto \sum_{j'm'_jm'_s} w_{j'm'_jm'_s}^{(\mu_2)} D_{(m'+\mu_2)m'_s}^{occ}(\varepsilon_{v^{-1}}) \tag{7}$$

Here $j$, $m_j$, $m_s$ and those with ' stand for the total angular momentum of the 2p core states, its z component and its spin, while $m$, and $m_s$ as well as those with ' denote the magnetic quantum number and the spin of the 3d states under consideration. Here $m = m_j - m_s$ is assumed for the 2p states. In the present calculation, the relativistic effects are considered for the $2p_{j=3/2}$ core states including the spin-orbit interaction for the states with $m_j = \pm 3/2$ and $\pm 1/2$, while the 3d spin is well defined by $m_s$. Furthermore, we take into account the Zeeman splitting of the 2p states due to the effective magnetic field of the 3d states. The weight coefficient $w_{jm_jm_s}^{(\mu)}$ is given by multiplication of the Clebsch-Goldan coeffcient and the Gaunt coefficient.

Finally we obtained the RIXS intensity as

$$\sigma^{\mu_1,\mu_2}(\nu_{in},\nu_{out}) \propto \sum_{jm_j}\sum_{m_s m'_s} \int d\varepsilon_{v^{-1}} \frac{w^{(\mu_2)}_{j'm'_j m'_s} w^{(\mu_1)}_{jm_j m_s} D^{\text{occ}}_{(m'+\mu_2)m'_s}(\varepsilon_{v^{-1}}) D^{\text{unocc}}_{(m+\mu_1)m_s}(\varepsilon_e)}{(\varepsilon_{v^{-1}} - E_{2p_{jm_j}} - h\nu_{out})^2 + \Gamma_i^2}$$

$$= \sum_{jm_j}\sum_{m_s m'_s} \int d\varepsilon_{v^{-1}} \frac{w^{(\mu_2)}_{j'm'_j m'_s} w^{(\mu_1)}_{jm_j m_s} D^{\text{occ}}_{(m'+\mu_2)m'_s}(\varepsilon_{v^{-1}}) D^{\text{unocc}}_{(m+\mu_1)m_s}(\varepsilon_{v^{-1}} + (h\nu_{in} - h\nu_{out}))}{(\varepsilon_{v^{-1}} - E_{2p_{jm_j}} - h\nu_{out})^2 + \Gamma_i^2} \quad (8).$$

In the simulation, we practically give the energy of unoccupied valence states $\varepsilon_e$ using the energy offset ($h\nu_1$) from Fermi energy ($E_F$) as $\varepsilon_e = E_F + h\nu_1$. Then, we further transform the equation (8) as

$$\sigma^{\mu_1,\mu_2}(\nu_1,\nu_{out}) \propto \sum_{jm_j}\sum_{m_s m'_s} \int d\varepsilon_{v^{-1}} \frac{w^{(\mu_2)}_{j'm'_j m'_s} w^{(\mu_1)}_{jm_j m_s} D^{\text{occ}}_{(m'+\mu_2)m'_s}(\varepsilon_{v^{-1}}) D^{\text{unocc}}_{(m+\mu_1)m_s}(h\nu_1 + E_F)}{(\varepsilon_{v^{-1}} - E_{2p_{jm_j}} - h\nu_{out})^2 + \Gamma_i^2} \quad (9)$$

In this model, the magnitude of the RIXS-MCD is proportional to the $D^{\text{unocc}}$ or the quantity of the core holes and the line shape reflects the energy distribution of the $D^{\text{occ}}$. Moreover, the spin polarization of the 2p core hole plays an essential role to the spin selective transition with dipole selection rule. Thus, the RIXS-MCD is clearly observed at the pre-edge of the $L_3$ XAS since the excitation from the spin-polarized $m_j = -3/2$ states is dominant. In the case of the $m_j = \pm 1/2$, however, the spins of the core hole are mixed, and the core hole spins can be relaxed between spin up and down before the fluorescence takes place. This allows the optical path in which the spins in the absorption process and emission process are different. The fluorescence spectra thus obtained are shown in Figs. 5(b) and 5(c). Note that the transition probability for the circular polarized photon was taken into account in the absorption process only, but it was averaged in the emission process since the polarization of the outgoing photon was not measured in the experiment.